\documentclass[12pt]{article}
\usepackage{graphicx}
\usepackage{dcolumn}
\usepackage{bm}
\usepackage{graphics,exscale}
\usepackage{amsmath}
\usepackage{amssymb}
\usepackage{amscd}
\usepackage{afterpage}
\usepackage{float,times}
\usepackage{subfigure}
\usepackage{rotating}
\usepackage{multirow}
\usepackage{fancyheadings}
\usepackage{epsfig}
\usepackage{theorem}
\usepackage{moreverb}
\usepackage{euscript}
\usepackage{psfrag}

\begin{document}
\pagestyle{empty}
{\hbox to\hsize{\hfill April 2008 }}

\vspace*{20mm}
\begin{center}

{\Large\bf Little Randall-Sundrum Model and a Multiply Warped Spacetime}\\
\vspace{1.0cm}

{\large Kristian L. McDonald\footnote{Email: klmcd@triumf.ca}}\\
\vspace{1.0cm}
{\it {Theory Group, TRIUMF, 4004 Wesbrook Mall, Vancouver, BC V6T2A3, Canada.}}
\vspace{.4cm}
\end{center}

\vspace{1cm}
\begin{abstract}
A recent work has investigated the possibility that the mass scale for the ultraviolet (UV) brane in the Randall-Sundrum (RS) model is of the order $10^3$~TeV. In this so called ``Little Randall-Sundrum'' (LRS) model the bounds on the gauge sector are less severe, permitting a lower Kaluza-Klein scale and cleaner discovery channels. However employing a low UV scale nullifies one major appeal of the RS model; namely the elegant explanation of the hierarchy between the Planck and weak scales. In this work we show that by localizing the gauge, fermion and scalar sector of the LRS model on a five dimensional slice of a doubly warped spacetime one may obtain the low UV brane scale employed in the LRS model and motivate the weak-Planck hierarchy. We also consider the generalization to an $n$-warped spacetime.
\end{abstract}

\vfill

\eject
\pagestyle{empty}
\setcounter{page}{1}
\setcounter{footnote}{0}
\pagestyle{plain}
\section{Introduction}
It is difficult to doubt the validity of the Standard Model (SM) of particle physics as an effective low energy description of nature given the body of precision electroweak data in agreement with SM predictions. Nonetheless the symmetry breaking structure of the SM remains experimentally unexplored and the possibility that non-SM physics may appear at the TeV scale persists. Our inability to understand the stability of the electroweak scale in the presence of ultraviolet (UV) sensitive quantum corrections (\emph{i.e.} the hierarchy problem) provides perhaps the biggest clue that new physics awaits us at TeV energies.

In its most severe form the hierarchy problem requires one to explain the stability of the electroweak scale relative to the Planck scale. One of the most promising approaches towards understanding this hierarchy involves the use of a warped or non-factorizable geometry~\cite{Randall:1999ee}. Assuming the metric 
\begin{eqnarray}
ds^2=e^{-k|y|}\eta_{\mu\nu}dx^\mu dx^\nu -dy^2\label{rs_metric},
\end{eqnarray}
one obtains an exponential hierarchy between the natural mass scale at the $y=0$ (UV) brane and the $y=L_y$ (IR) brane~\cite{Randall:1999ee}. Identifying these scales with the Planck and weak scales provides an elegant explanation for the weak/Planck hierarchy.

Unfortunately the simplest implementations of this idea are fraught with additional problems~\cite{Davoudiasl:1999jd}. Localizing fermions at the TeV brane results in a (little) hierarchy between the Kaluza-Klein (KK) and weak scales and thus the hierarchy problem is not completely resolved. If fermion\ \cite{Grossman:1999ra} and gauge~\cite{bulk_vector} degrees of freedom instead propagate in the five dimensional bulk, one must still extend the model to avoid conflict with oblique and non-oblique precision tests~\cite{Agashe:2003zs}. The observed absence of low energy violation of global lepton and baryon number charges also mandates an extension of the simplest proposals~\cite{Huber:2000ie}.
  
It is well known that precision electroweak data requires the SM cutoff to be $\Lambda_{UV}\gtrsim 10$~TeV. Thus even if one abandons attempts to immediately solve the hierarchy problem one must still explain this ``little hierarchy'' problem. Indeed the demands of the precision flavour data are even more severe and suggest that $\Lambda_{UV}\gtrsim10^2-10^3$~TeV, producing what has been called the ``weak-flavour hierarchy'' problem~\cite{Davoudiasl:2008hx}.

A recent work has investigated the possibility that a warped five dimensional geometry is responsible for the hierarchy between the weak flavour cutoff of $\sim10^3$~TeV and the weak scale~\cite{Davoudiasl:2008hx}. This so called ``Little Randall-Sundrum'' (LRS) model is a volume truncation of the usual RS model in which the natural mass scale of the UV brane is taken to be $\mathcal{O}(10^3)$~TeV, and is a candidate solution to the weak-flavour hierarchy problem. Interestingly many of the bounds on the gauge sector of the LRS model turn out to be less severe than their RS counterparts; corrections to the oblique $T$ parameter and the $Zb\bar{b}$ coupling both decrease as $kL_y$ decreases from its RS value, thereby permitting a lower KK scale. Furthermore additional (and perhaps more importantly, cleaner) discovery channels become viable for the LRS model as the signal to background ratio for, \emph{e.g.}, KK gauge bosons to decay to lepton pairs ($Z'\rightarrow \ell^+\ell^-$) increases significantly as $kL_y$ decreases. This greatly enhances the prospects for testing the model at the LHC.

Employing a low UV scale however unfortunately nullifies one of the major appeals of the RS model; namely the elegant explanation of the hierarchy between the Planck and weak scales. The natural exponential suppression of mass scales obtained with the non-factorizable geometry (\ref{rs_metric}) provides a powerful means by which to bridge the energy gap between the weak and Planck scales. In this work we show that by localizing the gauge, fermion and scalar sector of the LRS model on a five dimensional slice of a doubly warped spacetime one may obtain the low UV brane scale employed in the LRS model and employ a non-factorizable geometry to motivate the weak-Planck hierarchy.

We note that a number of authors have considered extended RS models with additional spacetime dimensions. The first work to consider a warped spacetime with two extra dimensions appears to be~\cite{Rubakov:1983bz}. An incomplete list of more recent studies includes~\cite{Kanti:2001vb,Multamaki:2002cq,Davoudiasl:2002wz,Collins:2001ni} and more specifically the concept of a doubly warped spacetime was investigated in~\cite{Choudhury:2006nj}. The doubly warped solutions to Einstein's equations obtained in the present work differ from those of~\cite{Choudhury:2006nj} as the present solutions permit large warping in both of the extra dimensions.

The layout of the paper is as follows. In Section~\ref{sec:2} we write down the metric and corresponding Einstein equations for a doubly warped spacetime. Section~\ref{sec:3} considers the realization of the LRS model in a doubly warped spacetime. The generalization to an $n$-warped spacetime is discussed in Section~\ref{sec:4} and we conclude in Section~\ref{sec:5}.
\section{Doubly Warped Spacetime\label{sec:2}}
Consider the following non-factorizable six dimensional spacetime metric:
\begin{eqnarray}
ds^2=a^2(y)b^2(v)\eta_{\mu\nu}dx^\mu dx^\nu -dy^2-dv^2\equiv G_{MN}dx^Mdx^N,
\end{eqnarray}
which corresponds to the usual RS1 metric augmented by an additional compact dimension in the limit $b(v)\rightarrow 1$. We take this additional dimension to be an $S^1/Z_2$ orbifold so that the fundamental space for the compact dimensions is $y\in[0,L_y]\equiv [y^{UV},y^{IR}]$ and $v\in[0,L_v]\equiv [v^{UV},v^{IR}]$.

The six dimensional Einstein equations with (4+1) branes localized at $y^{UV}$, $y^{IR}$, $v^{UV}$ and $v^{IR}$ are given by
\begin{eqnarray}
& &R_{MN}-\frac{1}{2}G_{MN}R^{(6)}=\nonumber\\
&-&\frac{1}{2\kappa}\left[\Lambda_{MN}+\frac{1}{\sqrt{-G_{55}}}\delta^{\tilde{\mu}}_M\delta^{\tilde{\nu}}_B\left\{G^{UV}_{\tilde{\mu}\tilde{\nu}}\delta(y)T_{y^{UV}N}^B+G^{IR}_{\tilde{\mu}\tilde{\nu}}\delta(y-L_y)T_{y^{IR}N}^B\right\}\right.\nonumber\\
& & \left.\qquad+\quad\frac{1}{\sqrt{-G_{66}}}\delta^{\bar{\mu}}_M\delta^{\bar{\nu}}_B\left\{G^{UV}_{\bar{\mu}\bar{\nu}}\delta(v)T_{v^{UV}N}^B+G^{IR}_{\bar{\mu}\bar{\nu}}\delta(v-L_v)T_{v^{IR}N}^B\right\}\right].
\end{eqnarray}
Here $\kappa=M_6^4/16\pi$ where $M_6$ is the six dimensional gravitational scale, the six dimensional indices are $M,N\in\{\mu,y,v\}$ and the five dimensional brane indices run over $\tilde{\mu},\tilde{\nu}\in\{\mu,v\}$ and $\bar{\mu},\bar{\nu}\in\{\mu,y\}$. The brane localized metrics are defined in terms of the bulk metric:
\begin{eqnarray}
& &G^{UV}_{\tilde{\mu}\tilde{\nu}}=G_{\tilde{\mu}\tilde{\nu}}(y=y^{UV})\quad ,\quad G^{IR}_{\tilde{\mu}\tilde{\nu}}=G_{\tilde{\mu}\tilde{\nu}}(y=y^{IR}),\nonumber\\
& &G^{UV}_{\bar{\mu}\bar{\nu}}=G_{\bar{\mu}\bar{\nu}}(v=v^{UV})\quad ,\quad G^{IR}_{\bar{\mu}\bar{\nu}}=G_{\bar{\mu}\bar{\nu}}(v=v^{IR}).
\end{eqnarray}
The most general bulk stress-energy tensor consistent with four dimensional Lorentz invariance has the inhomogeneous form
\begin{eqnarray}
\Lambda^M_N=\mathrm{diag}(\Lambda,\Lambda,\Lambda,\Lambda,\Lambda_5,\Lambda_6)
\end{eqnarray}
where either (or both) $\Lambda_5$ and $\Lambda_6$ may differ from $\Lambda$ (see~\cite{Kogan:2001yr,Multamaki:2002cq}). In what follows we refer to these quantities generically as the cosmological constant. The (4+1) brane tensions are
\begin{eqnarray}
T_{y^*N}^M&=&\mathrm{diag}(V_{y^*},V_{y^*},V_{y^*},V_{y^*},0,\bar{V}_{y^*}),\nonumber\\
 T_{v^*N}^M&=&\mathrm{diag}(V_{v^*},V_{v^*},V_{v^*},V_{v^*},\bar{V}_{v^*},0),
\end{eqnarray}
where $y^*\in \{y^{UV},y^{IR}\}$ and $v^*\in\{v^{UV},v^{IR}\}$. As discussed already in~\cite{Kogan:2001yr}, the inhomogeneities in $\Lambda^M_N$ and $T_{N}^M$ may result from distinct Casimir energy contributions in the different directions~\cite{Kogan:1983fp} or from a background three-form gauge field with a non-zero field strength~\cite{Chen:2000at}. Inhomogeneous brane tensions may also result from the zero point energies of brane localized fields~\cite{Chacko:1999eb}.

Considering first the bulk part of the Einstein equations gives
\begin{eqnarray}
3\left[\frac{a''}{a}+\frac{b''}{b}\right]+3\left[\frac{a'^2}{a^2}+\frac{b'^2}{b^2}\right]&=&-\frac{1}{2\kappa}\Lambda,\label{E1}\\
6\left[\frac{a'^2}{a^2}+\frac{b'^2}{b^2}\right] +4\frac{b''}{b}&=&-\frac{1}{2\kappa}\Lambda_5,\label{E2}\\
6\left[\frac{a'^2}{a^2}+\frac{b'^2}{b^2}\right] +4\frac{a''}{a}&=&-\frac{1}{2\kappa}\Lambda_6,\label{E3}
\end{eqnarray}
whose solutions are
\begin{eqnarray}
a(y)=\exp(-k_y|y|)\quad,\quad b(v)=\exp(-k_v|v|),
\end{eqnarray}
where
\begin{eqnarray}
k_y^2=\frac{1}{64\kappa}[3\Lambda_5-5\Lambda_6]\quad,\quad k_v^2=\frac{1}{64\kappa}[3\Lambda_6-5\Lambda_5],\label{6d_k_def}
\end{eqnarray}
and use of the orbifold reflection symmetries has been made. The Einstein equations also require the components of $\Lambda^M_N$ to obey 
\begin{eqnarray}
\frac{\Lambda}{6}-\frac{1}{16}[\Lambda_5+\Lambda_6]=0.
\end{eqnarray}
 Including the branes gives the following jump equations,
\begin{eqnarray}
-4\left.\frac{[a']}{a}\right|_{y^*}=\frac{1}{2\kappa}\bar{V}_{y^*}\quad,\quad -4\left.\frac{[b']}{b}\right|_{v^*}=\frac{1}{2\kappa}\bar{V}_{v^*},
\end{eqnarray}
and equation (\ref{E1}) requires $\bar{V}_{y^*}=4V_{y^*}/3$ and $\bar{V}_{v^*}=4V_{v^*}/3$ (a result found also in~\cite{Kanti:2001vb}). As per usual in RS models, the jump conditions lead to a fine tuning between the bulk cosmological constant(s) and the brane tensions:
\begin{eqnarray}
\bar{V}_{y_{UV}}&=&-\bar{V}_{y_{IR}}=\frac{1}{4k_y}(\Lambda_5-\frac{5}{3}\Lambda_6),\nonumber\\
\bar{V}_{v_{UV}}&=&-\bar{V}_{v_{IR}}=\frac{1}{4k_v}(\Lambda_6-\frac{5}{3}\Lambda_5),
\end{eqnarray}
and the four dimensional Planck scale may be expressed in terms of the fundamental gravity scale $M_6$:
\begin{eqnarray}
M_{Pl}^2=\frac{M_6^4}{k_yk_v}\left\{1-e^{-2k_yL_y}\right\}\left\{1-e^{-2k_vL_v}\right\}.
\end{eqnarray}
\section{The Little Randall-Sundrum Model\label{sec:3}}
Now let us consider the gauge, Higgs and fermion sector of the LRS model to be localized on the (4+1) brane at $v=L_v$ and denote the Lagrangian describing these fields as $\mathcal{L}_{LRS}$. The relevant action is 
\begin{eqnarray}
S_6=\int d^6x \sqrt{-G^{v_{IR}}}\mathcal{L}_{LRS}\delta(v-L_v).
\end{eqnarray}
After integrating out the sixth dimension the $v$ dependent warp factors which arise from the metric can be absorbed into a wavefunction renormalization of the field operators. The only place where the $v$ space warp factor is not removed by this process is in the Higgs potential. Taking this to be localized at the IR (3+1) brane of the five dimensional slice $v=L_v$ gives
\begin{eqnarray}
S_6^H&=& \int d^6x\sqrt{-\tilde{G}}\left\{\tilde{G}^{\mu\nu}D_\mu HD_\nu H - \lambda(H^2-\bar{v}_{0}^2)\right\}\delta(v-L_v)\delta(y-L_y) +....\nonumber\\
&=& \int d^5x\sqrt{\tilde{g}}\left\{\tilde{g}^{\mu\nu}D_\mu HD_\nu H - \lambda(H^2-v_{0}^2)\right\}\delta(y-L_y) +....
\end{eqnarray}
where $H$ has been renormalized and $v_0=\bar{v}_0\exp(-k_vL_v)$. Here $\tilde{G}_{\mu\nu}$ is the 4D metric at $v,y=L_{v,y}$ and the rescaled 4D metric is $\tilde{g}_{\mu\nu}=b^{-2}(v_{IR})\tilde{G}_{\mu\nu}$. From the vantage point of the effective five dimensional theory it appears that the natural mass scale for the dimensional input parameter in the Higgs potential is $v_0$. This will also appear to be the natural mass scale for the $UV$ brane in the effective five dimensional theory. Consequently if the fundamental Higgs mass parameter $\bar{v}_0$ takes its natural value of order $M_6$, the high energy (UV brane) scale in the effective five dimensional theory is warped down to $\sim\exp(-k_vL_v)M_6$. This motivates the use of a low UV brane scale (compared to the Planck scale) in RS1 models and in particular may motivate the order $10^3$~TeV UV scale advocated in the LRS model.

To obtain the electroweak scale one integrates out the $y$ space:
\begin{eqnarray}
S_6^H&=& \int d^4x\sqrt{-g}\left\{g^{\mu\nu}D_\mu HD_\nu H - \lambda(H^2-v_{EW}^2)\right\} +....
\end{eqnarray}
where $g$ denotes the 4D metric and the $y$ space warp factor has been absorbed by the wavefunction renormalization $H\rightarrow e^{k_yL_y}H$. The electroweak symmetry breaking scale is thus
\begin{eqnarray}
v_{EW}\equiv \exp(-k_yL_y)v_0=\exp(-k_yL_y-k_vL_v)\bar{v}_0,
\end{eqnarray}
making obvious the double warping which reduces the Planck scale sized input parameter $\bar{v}_0$ to the electroweak scale $v_{EW}$. The LRS scenario is realized on the (4+1) brane at $v_{IR}$ for $e^{-k_vL_v}\approx 10^{-13}$ so that $k_vL_v/k_yL_y\approx 4.3$ and one requires, for example, $k_v\approx 4.3 k_y$ ($2.1 k_y$) for $L_v=L_y$ ($2.1 L_y$). We see that only a very mild hierarchy between the warping parameters of the extra dimensions is necessary to achieve an LRS like setup.

With the SM gauge and fermion sectors localized at $v=L_v$ it is clear that the gauge interactions of the present model will, by construction, match those of~\cite{Davoudiasl:2008hx} and the gauge and flavour bounds obtained in~\cite{Davoudiasl:2008hx} will remain valid. One may wonder how the two models would thus be discernible. One key difference is the KK structure of the gravitational fluctuations; the present model possesses additional states relative to the five dimensional model due to the quantized graviton momenta in the sixth dimension (see for example~\cite{Davoudiasl:2002wz}). Thus if LRS phenomenology were to be observed one could experimentally study the KK structure of the graviton to determine if nature employs a doubly warped spacetime.

\section{n-Warped Spacetime\label{sec:4}}
We have shown that a doubly warped spacetime may generate the hierarchies between the weak scale, the Planck scale and an intermediate scale and that the intermediate scale may be interpreted as the weak-flavour scale employed in the LRS model. It is possible however that nature employs a number of intermediate scales between $v_{EW}$ and $M_{Pl}$; another well studied candidate scale being the Grand Unified Theory (GUT) scale. In this section we consider the generalization of a doubly warped spacetime to an $n$-warped spacetime. This permits the warping of Planck scale input parameters to generate $n-1$ intermediate scales between the weak and Planck scales. Such a structure may be interesting for model building purposes when trying, for example, to embed the LRS model into a GUT framework.

We consider the $(4+n)$ dimensional spacetime with metric
\begin{eqnarray}
ds^2=\left(\prod_{i=1}^{n}a_i^2(y_i)\right)\eta_{\mu\nu}dx^\mu dx^\nu -\sum_{i=1}^{n}dy
_i^2\equiv G_{MN}dx^Mdx^N
\end{eqnarray}
where the indices $M,N$ label the $(4+n)$ dimensional spacetime, $N,M\in (\mu,y_1,y_2....y_n)$, and $y_i$ labels the $i$th compact extra dimension ($i=1,2,...n$) which we take to be an $S^1/Z_2$ orbifold. The fundamental space for a given compact dimension is $y_i\in[0,L_i]\equiv [y^{UV}_i,y^{IR}_i]$ and we will use $y^*_i$ to denote either of the boundary points $y^{UV}_i$ or $y^{IR}_i$.

The full $(4+n)$ dimensional Einstein equations are
\begin{eqnarray}
\mathcal{G}_{MN}&\equiv& R_{MN}-\frac{1}{2}G_{MN}R^{(4+n)}=-\left.\frac{1}{2\kappa}\right[\Lambda_{MN}\nonumber\\
&+&\left.\sum_{i=1}^n\frac{\delta^{\mu_i}_M\delta^{\nu_i}_B}{\sqrt{-G_{y_iy_i}}}\left\{G^{y_i^{UV}}_{\mu_i\nu_i}\delta(y_i-y^{UV}_i)T_{y_i^{UV}N}^B+G^{y_i^{IR}}_{\mu_i\nu_i}\delta(y_i-y^{IR}_i)T_{y_i^{IR}N}^B\right\}\right].
\end{eqnarray}
Here $\kappa=M_{(4+n)}^{2+n}/16\pi$ and $\mathcal{G}_{MN}$ ($M_{(4+n)}$) is the generalized $(4+n)$ dimensional Einstein tensor (gravitational scale). The brane coordinates $x^{\mu_i}$ run over all coordinates except $y_i$ so that $x^{\mu_i},x^{\nu_i}\in\{x^\mu,y_1,...,y_{i-1},y_{i+1},...,y_n\}$. The cosmological constant is
\begin{eqnarray}
\Lambda^M_N=\mathrm{diag}(\Lambda,\Lambda,\Lambda,\Lambda,\Lambda_{y_1},\Lambda_{y_{2}},....,\Lambda_{y_n}),
\end{eqnarray}
and $T^M_{y_i^*N}$ is the brane tension of the $4+(n-1)$ brane localized at $y_i^*$ with metric 
\begin{eqnarray}
G^{y_i^*}_{\mu_i\nu_i}=G_{\mu_i\nu_i}(y_i=y_i^*).
\end{eqnarray}
Considering first the bulk part of the Einstein equations, the $\mathcal{G}_{\mu\nu}$ equations give
\begin{eqnarray}
\sum_{i=1}^n3\left\{\frac{a'^2_i}{a_i^2}+\frac{a''_i}{a_i}\right\} = -\frac{1}{2\kappa}\Lambda
\end{eqnarray}
and the $n$ equations from the $\mathcal{G}_{y_iy_i}$ components are
\begin{eqnarray}
\sum_{j=1}^n\left\{6\frac{a'^2_j}{a_j^2}+4\frac{a''_j}{a_j}(1-\delta_{ij})\right\}=-\frac{1}{2\kappa}\Lambda_{y_i}.\label{Eii_4+n}
\end{eqnarray}
The solutions to these equations are exponentials
\begin{eqnarray}
a_i(y_i)=\exp(-k_i|y_i|),\quad \forall \ i.
\end{eqnarray}
Analogous to the six dimensional expressions (\ref{6d_k_def}), the $n$ relations (\ref{Eii_4+n}) can be used to express the $n$ constants $k_i$ in terms of the constants $\Lambda_{y_i}$:
\begin{eqnarray}
k_i^2=\frac{1}{8\kappa}\frac{1}{(5n-2)}\left\{(5n-7)\Lambda_{y_i} -5\sum_{j\ne i}\Lambda_{y_j}\right\}.
\end{eqnarray}
The $\mathcal{G}_{\mu\nu}$ equation then relates the bulk cosmological constants
\begin{eqnarray}
\frac{\Lambda}{6}-\sum_{i=1}^{n}\frac{\Lambda_{y_i}}{(10n-4)}=0,
\end{eqnarray}
and we note that $\Lambda=\Lambda_{y_1}$ for $n=1$ as found in the RS1 model. Including the localized brane tensions on the right hand side of the Einstein equations gives the jump conditions for the derivatives of the functions $a_i$ at the points $y_i^*$. Using the Einstein equations one finds that the brane tension for the $4+(n-1)$ brane localized at $y_i^*$ has the form
\begin{eqnarray}
T_{y^*_iN}^M=\delta^M_\mu\delta_N^\mu V_{y^*_i} + \delta^M_{y_j}\delta_N^{y_j} \bar{V}_{y^*_i}(1-\delta_{y_iy_j}),
\end{eqnarray}
where no summation is implied by the repeated indices $\mu$ and $y_j$ and the zero entry occurs in the $M=N=y_i$ element, $T_{y^*_iy_i}^{y_i}=0$. The jump equations are thus
\begin{eqnarray}
-4\left.\frac{[a'_i]}{a_i}\right|_{y_i^*}=\frac{1}{2\kappa}\bar{V}_{y^*_i},
\end{eqnarray}
with the Einstein equations requiring $\bar{V}_{y^*_i}=4V_{y^*_i}/3$ and the usual bulk-brane cosmological constant fine tunings:
\begin{eqnarray}
\bar{V}_{y_i^{UV}}=-\bar{V}_{y^{IR}_i}=\frac{2}{k_i}\frac{1}{(5n-2)}\left\{(5n-7)\Lambda_{y_i} -5\sum_{j\ne i}\Lambda_{y_j}\right\}.
\end{eqnarray}
Finally the four dimensional Planck scale may be expressed in terms of the $(4+n)$ dimensional fundamental scale $M_{(4+n)}$ as
\begin{eqnarray}
M_{Pl}^2=M_{(4+n)}^{2+n}\prod_{i=1}^n\frac{1}{k_i}\left\{1-e^{-2k_iL_i}\right\}.
\end{eqnarray}
To see the effect of the multiple warping we may consider the SM Higgs doublet to be localized at the $(3+1)$ brane defined by  $y_{i}^{IR\ \ }\forall \ i$. Thus the Lagrangian for $H$ is
\begin{eqnarray}
S_{(4+n)}^H&=& \int d^{4+n}x\sqrt{-G}\left\{G^{\mu\nu}D_\mu HD_\nu H - \lambda(H^2-\bar{v}_{0}^2)\right\}\prod_{i=1}^n\delta(y_i-y^{IR}_i)\nonumber\\
&=& \int d^4x\sqrt{-g}\left\{g^{\mu\nu}D_\mu HD_\nu H - \lambda(H^2-v_{EW}^2)\right\},
\end{eqnarray}
where we have integrated out the extra dimensions, performed the wavefunction renormalization $H\rightarrow \left(\prod_ie^{k_iL_i}\right)H$ and defined the metric on the Higgs $(3+1)$ brane as $g$. The electroweak scale is thus
\begin{eqnarray}
v_{EW}=\left(\prod_{i=1}^ne^{-k_iL_i}\right)\bar{v}_0.
\end{eqnarray}
More generally the vacuum expectation value of a scalar localized on a $(3+1)$ brane at the IR end of $m<n$ extra dimensions and the UV end of $(n-m)$ dimensions will experience an $m$-fold warping given by
 \begin{eqnarray}
v_{m}=\left(\prod_{i=1}^me^{-k_iL_i}\right)\bar{v}_0,
\end{eqnarray}
where we have ordered the coordinates $y_i$ such the the dimensions of IR localization occur first. Thus, as expected, if the SM Higgs is localized at the $(3+1)$ brane $y_{i}^{IR\ \ }\forall \ i$ one may generate $(n-1)$ intermediate scales between the electroweak scale and the fundamental gravitational scale with an $n$-warped spacetime. In this way nature could, for example, generate both the GUT scale and the weak scale from Planck scale input parameters\footnote{The notion of a partial or mini-warping from the Planck scale to the GUT scale has been examined in~\cite{Fukuyama:2007ph}.}.
\section{Conclusion\label{sec:5}}
We have presented a solution to Einstein's equations in six dimensions with two warped compact extra dimensions (a doubly warped spacetime). This allows one to employ spacetime warping to explain the large hierarchy between the Planck and the weak scales whilst considering SM gauge phenomenology on a truncated RS1 spacetime (as in the LRS model). We have also generalized the concept of a doubly warped spacetime to an $n$-warped spacetime and shown that more complicated mass scale hierarchies may result. Important open issues in this framework include the nature of radius stabilization and the structure of the graviton KK tower.
\section*{Acknowledgements}
The author is grateful to D. P. George, J. N. Ng and R. R. Volkas for helpful communications regarding the manuscript.

\end{document}